\documentclass{aa}
\usepackage{graphicx}

\long\def\jumpover#1{{}}
\def \ngth{\negthinspace}
\def \th{\thinspace}
\def \viz{{\it viz.}}
\def \ie{{\it i.e.}}
\def\ah{{\scriptscriptstyle 1/2}}
\def \Teff{{$T_{\rm ef\!f} $}}

\begin{document}


\title{Nonlinear Beat Cepheid and RR Lyrae Models}
\titlerunning{Nonlinear Beat Models}
\authorrunning{Koll\'ath et al.}
\author{Z. Koll\'ath\inst{1} \and J. R. Buchler\inst{2} \and 
R. Szab\'o\inst{1} \and Z. Csubry\inst{1}}
\institute{
Konkoly Observatory, Budapest, HUNGARY\\
\email{kollath@konkoly.hu}
\and
Physics Department, University of Florida, Gainesville, FL 32611, USA\\
\email{buchler@phys.ufl.edu}
}

\offprints{Z. Koll\'ath}
\date{submitted Ocober 2, 2001}

\abstract{
The numerical hydrodynamic modelling of beat Cepheid behavior has been a
long standing quest in which purely radiative models had failed consistently.
We find that beat pulsations occur naturally when {\it turbulent convection} is
accounted for in our hydrodynamics code.  The development of a relaxation code
and of a Floquet stability analysis greatly facilitates the search for and the
analysis of beat Cepheid models.  The conditions for the occurrence of beat
behavior can be understood easily and at a fundamental level with the help of
amplitude equations.
}

\maketitle

\section{Introduction}

Double-mode or beat pulsations are not uncommon among RR~Lyrae (the RRd stars)
and classical Cepheids ({\it cf.\ } {{\it e.g.}} Kov\'acs 2001).  Yet the
numerical modelling of this phenomenon remained an unmet challenge for years.
Partial success was achieved with purely radiative codes (Kov\'acs \& Buchler
1993), but it took the incorporation of a model for turbulent convection into
the hydrocodes to obtain results that are both robust and have astrophysical
parameters that are in agreement with the observations.  Essentially
simultaneously, but independently, Koll\'ath { \it et al.} (1998) could model
double-mode pulsations in Cepheid models and Feuchtinger (1998) in RR~Lyrae
models. In this paper we present an extension of this earlier work.
We examine in some detail specific examples of both RR Lyrae and Cepheid
models, as well as sequences of such models (in which L and M are held fixed
and {$T_{\rm {ef\!f}} $}\ is varied, thus mimicking approximately the
horizontal branch and the Cepheid loops, respectively).

\section{Turbulent Convection Model Equations}

The turbulent convection model equations that we have incorporated in our
hydrodynamics code are essentially those of Gehmeyr \& Winkler (1992) and
Kuhfu\ss~(1986). They differ in the expression for the convective flux $F_c$
from those of Stellingwerf (1982) and Bono \& Stellingwerf (1994) and those of
original paper (Yecko, Koll\'ath \& Buchler 1998), but are essentially the same
as used by Wuchterl \& Feuchtinger (1998).  To be specific we reproduce here
the equations that we use in our hydrodynamics code.

The fluid dynamics part of the model calculations are given by the
following equations:
 \begin{eqnarray}
  {du\over dt} &=& -{1\over\rho}{\partial \over\partial r}
\left(p+\th\th p_t+p_\nu\th\th \right)
   - {G M_r\over r^2} \quad\quad \\
 & & \nonumber \\
 \ \ \ \ {d\th e\over dt} +p\th  {d\th v\over dt}
 &=& -{1\over\rho r^2} {\partial \over\partial r} \left[ r^2
\left(F_r+\th\th F_c\th\th \right)\right] 
 -\th \mathbf{{\cal C}}
 \quad\quad
 \end{eqnarray}
The turbulent motion of the gas and the convection interacts with the
hydrodynamics of the radial motion through the convective flux $F_c$, the
viscous eddy pressure $p_\nu$ the turbulent pressure $p_t$, and finally,
through an energy coupling term ${\cal C}$.  The turbulent energy $e_t$ is
determined by a time dependent diffusion equation
 \begin{equation}
   {de_t\over dt} +
    \left(p_t+p_\nu\right)\th  {d\th v\over dt}
  = -{1\over\rho r^2} {\partial \over\partial r}\left( r^2 F_t\right)
   +{\cal C} \quad\quad \nonumber\\
\end{equation}
The coupling term that connects the gas and the turbulent energy
equations is given by:
 \begin{equation}
 {\cal C}=  \alpha_d\th { e^{\scriptscriptstyle 1/2}_t\over\Lambda}
  \left(S_t - e_t\right),
 \end{equation}
where $\Lambda = \alpha_\Lambda\th H_p$, $H_p = p\th
r^{\scriptstyle 2}/(\rho GM)$ is the pressure scale height,
$\alpha_\Lambda$ is the mixing length parameter.  The
$\alpha_d$'s are dimensionless parameters of order unity.
Both the convective flux and the source term of the turbulent energy ($S_t$)
depend on the dimensionless entropy gradient 

\begin{eqnarray}
Y   &=& -{H_p/c_p} \th\th {\partial s/ \partial r} \\
S_t &=& (\alpha_s\th \alpha_\Lambda)^{\scriptstyle 2} 
      {p\over \rho} \th\beta  T\th\th Y \, f_{pec}
\end{eqnarray}
The remaining quantities are defined as
\begin{eqnarray}
  p_t   &=& \alpha_p \th \rho \th e_t ,\\
  p_\nu &=& - \frac{4} {3} \alpha_\nu\th \rho \Lambda   
                  e^{\scriptscriptstyle 1/2}_t
                  r{\partial\th \over\partial r}{u\over r} \\
  F_t   &=& -\alpha_t\th \rho \Lambda  e^{\scriptscriptstyle 1/2}_t 
                      {\partial\th\th e_t 
                      \over\partial r} \\
  F_c   &=& \alpha_c\alpha_\Lambda\th \rho 
             e_t^{{\scriptscriptstyle 1/2}} \th
             c_p T\th  Y \, f_{pec}.
\end{eqnarray}
\begin{eqnarray}
  f_{pec}   = {1\over 1+\alpha_r P  e^{-1}},\quad\quad 
            &Pe& =  D_c/D_r  \\
       D_r = {4\over3} {a c T^3 \over \kappa\rho^2 c_p},\quad\quad
       &D_c& = \Lambda e_t^{\scriptscriptstyle 1/2}
\end{eqnarray}
The turbulent convective equations can be derived on dimensional and physical
grounds with a number of dimensionless parameters ($\alpha$'s) that are of
order unity but for which theory provides no numerical values.  Our approach
has been to attempt to calibrate them through a comparison of our results with
the observational data.

The calibration of the $\alpha$'s is a daunting task because of the large
number of parameters on the one hand, and the large number of observational
constraints, many of which require extensive full amplitude hydrodynamical
model calculations.  For example, important constraints are the locations and
widths of the fundamental and first overtone instability strips, the Fourier
decomposition coefficients of the light curve and of the radial velocity data,
with their dependence on metallicity.  Such a global calibration has not yet
been performed, and it is not even sure that it possible given the rather
simple turbulent convection recipe that we use.  In that sense our results
must be regarded as tentative.  We stress however, that the types of behavior
that we describe here are not sensitive to these $\alpha$'s, but that the
masses, luminosities and {$T_{\rm {ef\!f}} $}'s at which they occur depend on
them.  In Koll\'ath \& Buchler (2001) we have presented typical modal selection
diagrams for Cepheid and RR~Lyrae models (figures 14 and 15) that show the
calculated pulsational behavior of the models in a Hertzsprung-Russell diagram.

\section{Transient Behavior of Models}

The first issue that we address concerns the confirmation that a given model is
indeed undergoing stable double-mode pulsations rather than being involved in a
switch from one mode to another.
Deciding on the basis of a single hydrodynamical calculation whether a model is
undergoing steady double-mode pulsations can be fraught with peril.  The model
may give the false impression of having achieved steady behavior when it is
actually still in a transient state even after thousands of periods of
integration, and it will end up pulsating with a single frequency.  The reason
for this will become clear shortly.

\begin{figure*}
\resizebox{18cm}{!}{\includegraphics{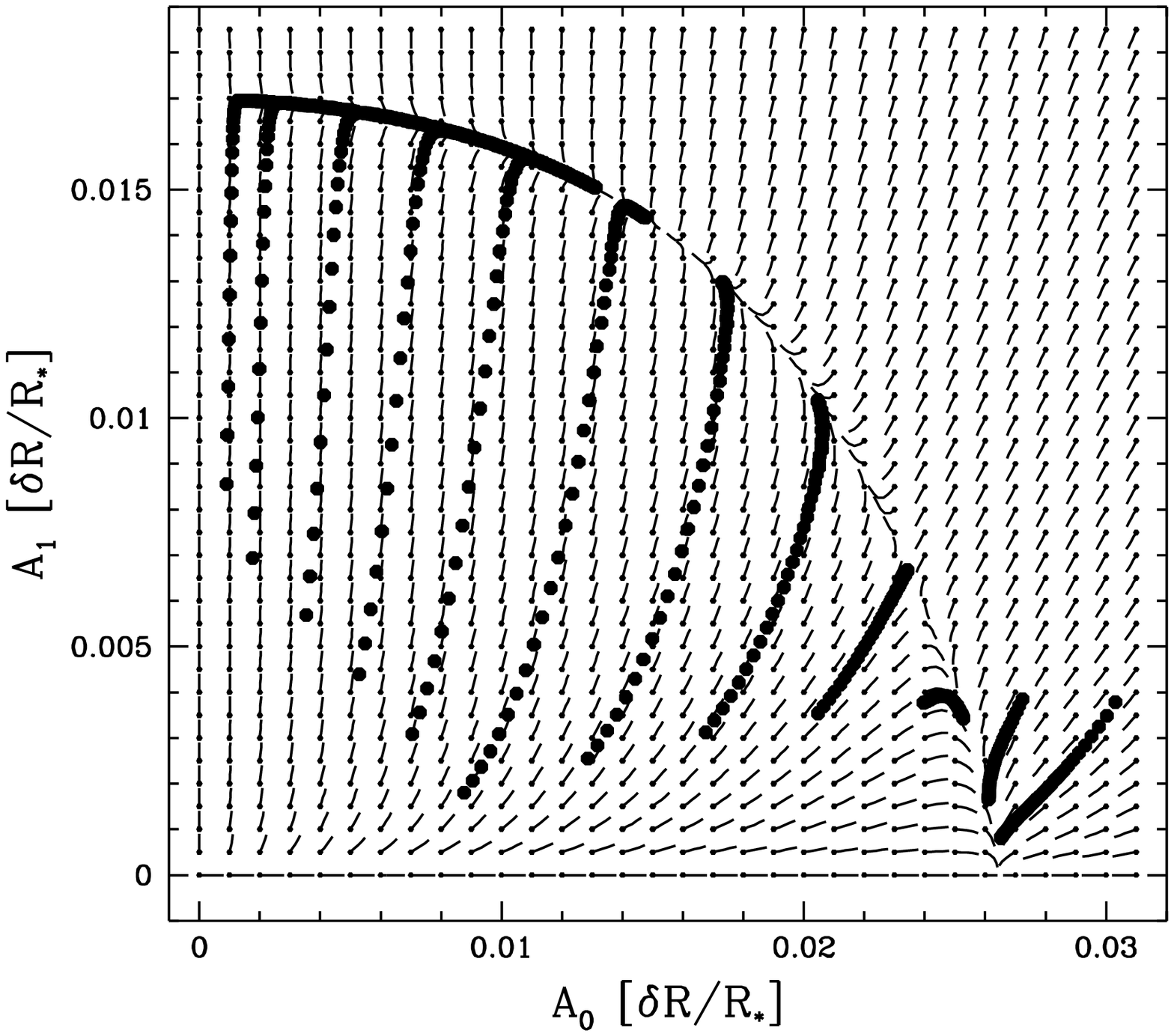}
                    \includegraphics{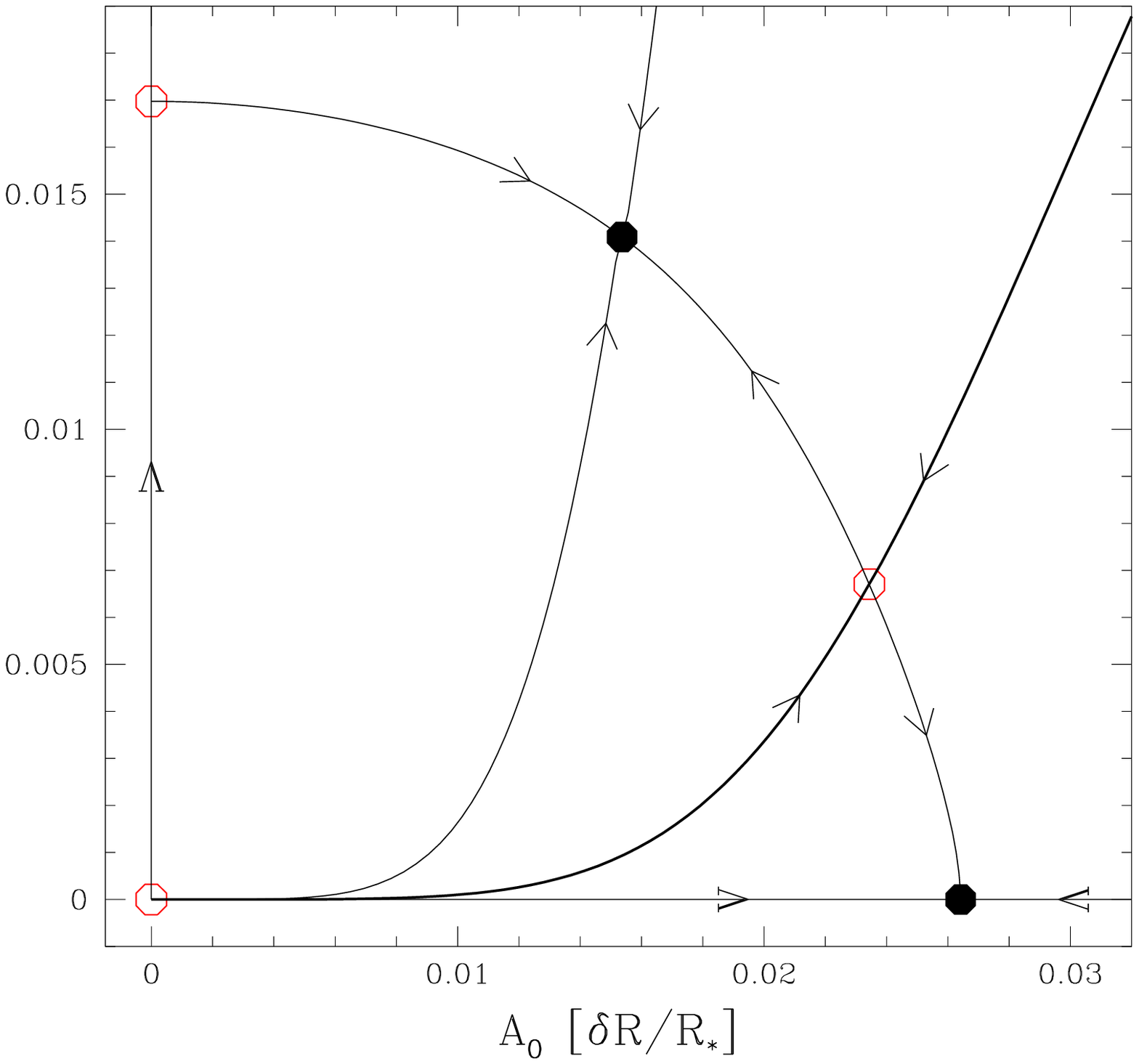}}
\caption{Cepheid model with {$T_{\rm {ef\!f}} $}=5580\th K.\hfill\break
{\sl Left}: The flow in the
($A_0,A_1$) phase space. The large dots represent the hydrodynamical results
at equal time intervals, so that the spacing of the dots gives an indication of
the speed.  The short lines represents the flow field (dots denote the base of
the normalized vectors).\hfill\break
{\sl Right}: The stable/unstable fixed points are denoted by filled/open circles.
The integral lines (heteroclinic connections) that connect the fixed points are
shown as solid lines with arrows indicating the evolution direction.  The thick
line is also the separatrix, {\it i.e.} the boundary between the basins of
attraction of the two stable fixed points, \viz the stable DM 
and the fundamental. }
\label{figflow}
\end{figure*}

\subsection{Time Dependence of Amplitudes and Phases}

One of the several tools that are needed is the calculation of the time
dependence of the amplitudes and phases of the modes that are excited in the
models.  Instead of using a time dependent Fourier analysis as advocated by
Kov\'acs, Buchler \& Davis (1987, KBD87) we now use the much more powerful and
accurate {\it analytical signal method} of G\'abor (1946) \footnote{Physicists
are generally familiar with this analytic signal concept through the
Kramers--Kronig dispersion relations (Jackson 1975).  If $s(t)$ represents the
real part of an assumed complex analytical function $Z(t)$, then the imaginary
part $\tilde s(t)$ of $Z(t)$ can be obtained via a Cauchy integral, which
through contour deformation becomes a Hilbert transform, which in turn can be
converted into a one-sided Fourier transform.} to reconstruct the time
dependent amplitudes and frequencies of the pulsation modes.  ({\it cf.\ }
{{\it e.g.},\ } Cohen 1994).  The method can be extended to multi-component
signals with the help of filters that restrict the power to the desired
frequency components.  It is convenient to make this filtering in Fourier space
where it involves just a product, and to combine it with the definition of the
analytical signal (Koll\'ath \& Buchler 2001)
 \begin{eqnarray}
   \ngth\ngth
  Z_{k}(t) & = & a_{k}(t)e^{i\varphi_{k}(t)}  
  \nonumber 
  \\
           & = & \frac{1}{\pi}\int^{\infty}_{0} d\omega 
               H(\omega -\omega _{k})e^{i\omega t}
              \int_{-\infty}^\infty dt' s(t')e^{-i\omega t'},
\end{eqnarray}
where $s(t)$ is the real valued signal that is to be analyzed, $H(\omega
-\omega _{k}) $ is the filtering window that is centered on the desired
$\omega_k$.  A Gaussian window with a half width of $\omega _{k}/10$
generally gives
satisfactory results for the stellar pulsations.  The resulting amplitudes and
phases describe then the transient behavior of the given mode.  The analytical
signal approach yields a much better resolution than the KBD87 approach because
it is not necessary to average the amplitudes in time to get smooth results.

With the analytical signal method it is therefore very easy and fast to define
unambiguously the instantaneous phase $\varphi(t)$ and amplitude $A(t)$ of a
signal.

\subsection{Phase Plots}

The computed time dependent amplitudes, in our case $A_0(t)$ and $A_1(t)$, can
then be used to visualize the transient behavior of the pulsating model in an
($A_0$,$A_1$) phase space.  Our amplitudes are defined as those of the
normalized radial displacement of the photosphere, $\delta R_{ph}/R_*$, where
$R_*$ is the photospheric radius of the equilibrium model.

In Fig.~\ref{figflow} on the left we display
the results of a number of hydrodynamical calculations made for the same
Galactic Cepheid model (M=4.7~{$M_{\sun} $}, L=1335~{$L_{\sun} $}, {$T_{\rm
{ef\!f}} $} = 5580~K, Z=0.02), but started (kicked from the unstable
equilibrium) with different initial conditions.  The tracks all start with
small amplitudes (lower left side).

Several notable features stand out in the left of Fig.~\ref{figflow}.  Since
the thick dots denote points that are equally separated in time one sees that
the models all evolve very quickly toward an arc along which the transient
evolution is extremely slow.  If only one model were computed this
'convergence' could easily be misinterpreted as a steady double-mode pulsation.
The figure indicates that the point on the $A_0$ axis with ($A_0= 0.026,
A_1=0$) is an attractor.  The three rightmost tracks transit to this
fundamental attractor after a sufficiently long integration time.  This
corresponds of course to the fundamental limit cycle.  Similarly the point on
the arc with ($A_0=0.023, A_1=0.007$) repels the tracks, and it corresponds to
an unstable double-mode pulsation.  Further up along the arc a second attractor
with ($A_0=0.015, A_1=0.014$) is clearly visible.  Six of the tracks converge
toward it from the left and three of them from the right.  This attractor
corresponds to stable double-mode pulsations.  Finally, one notes that the
single-mode overtone limit cycle at ($A_0=0, A_1=0.017$) repels because it is
unstable.  All tracks run away from the origin because the equilibrium model is
linearly unstable in both modes.

We shall see that this whole picture can be clearly understood after we have
introduced amplitude equations.  They allow us to draw the flowfield that has
been superposed on the left-hand figure.

\begin{figure*}
\resizebox{18cm}{!}{\includegraphics{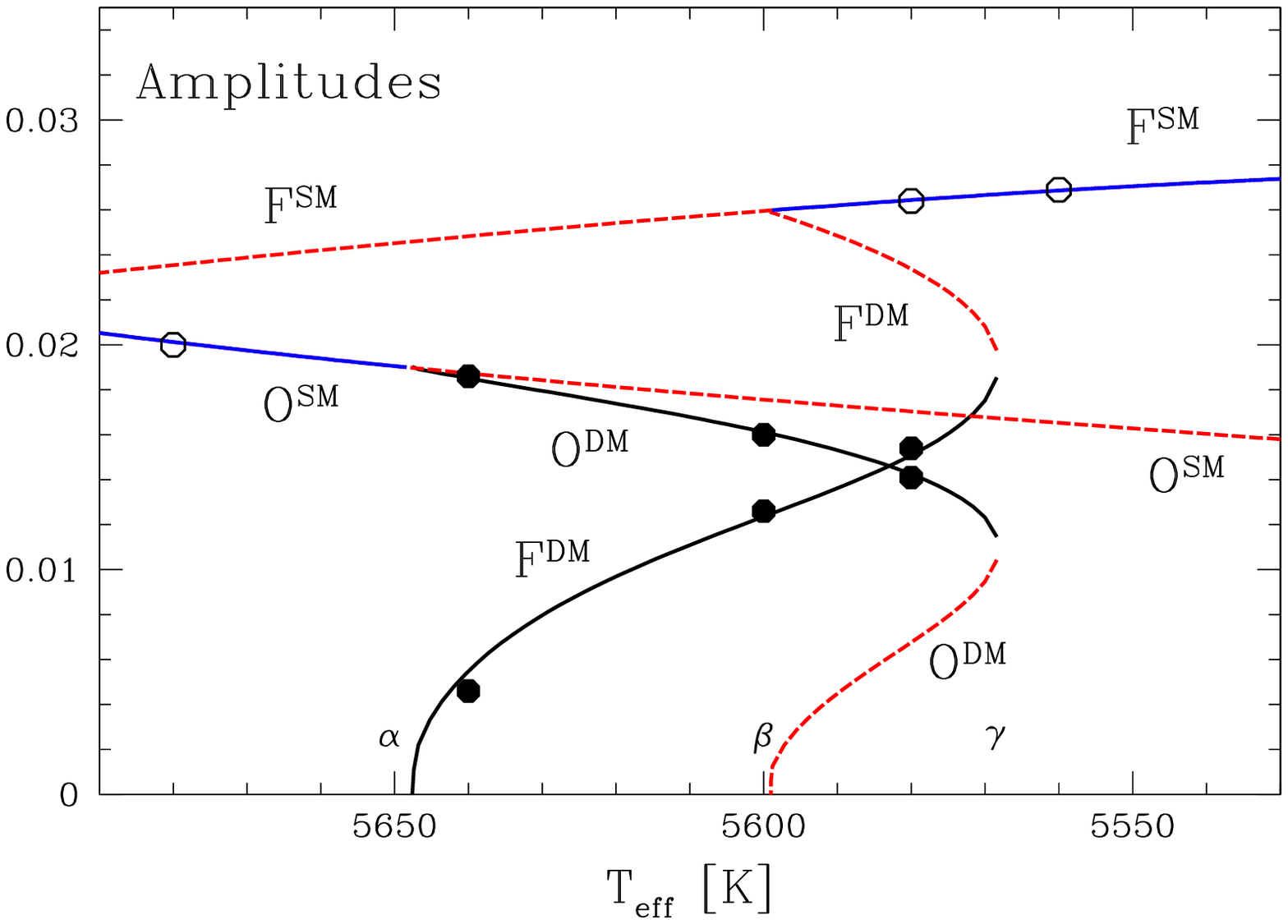}
                    \includegraphics{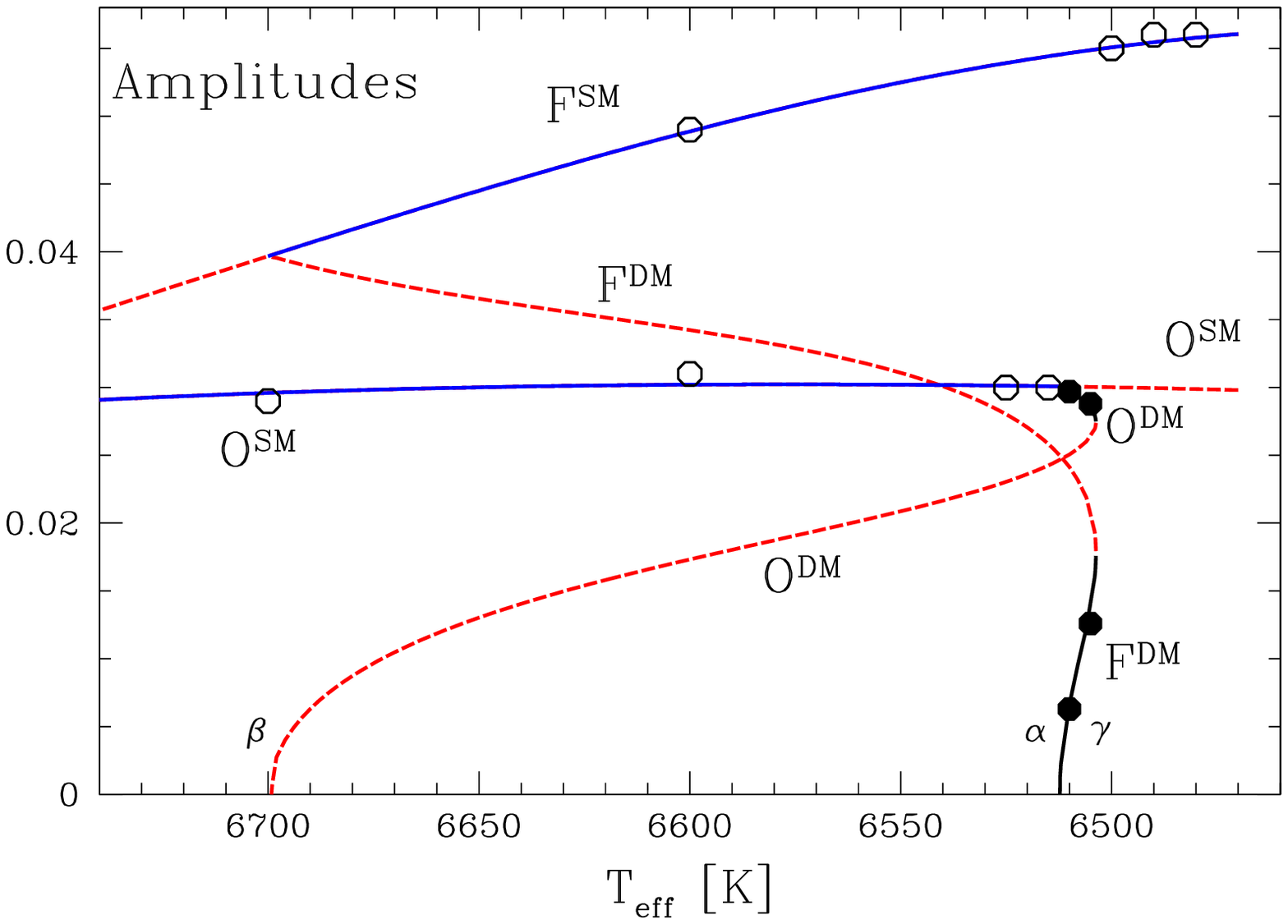}}
\caption{
{\sl Left}: Behavior of the amplitudes 
of single-mode (SM) and double-mode (DM) pulsations
along a constant $L$ convective Cepheid model sequence; 
Solid lines: stable pulsations, 
dashed lines: unstable pulsations; {\it cf.\ } 
text. The filled and open circles locate calculated hydro-models.\hfill\break
{\sl Right}: RR~Lyrae sequence. (Amplitude in units of $\delta R_{ph}/R_*$.)
}
\label{figfpae}
\end{figure*}

\begin{figure*}
\begin{center}
\vskip 1cm
\resizebox{17cm}{!}{\includegraphics{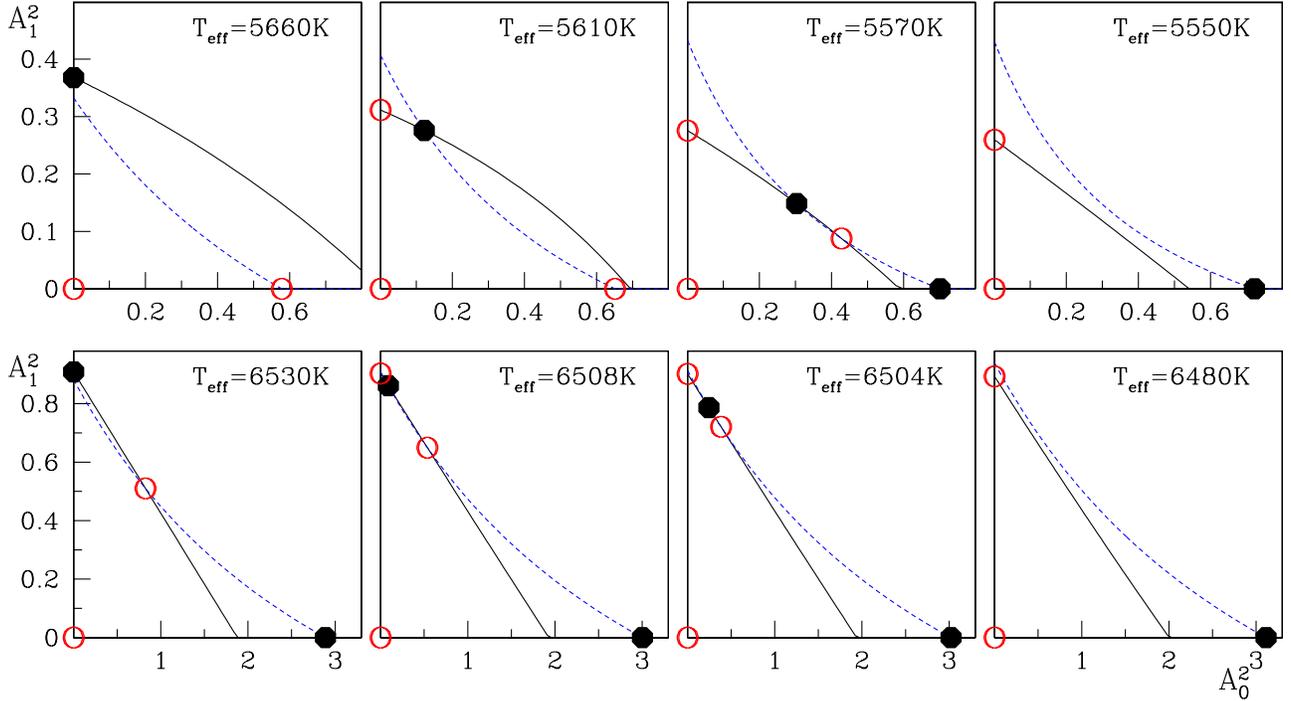}}
\caption{
Switch from single-mode overtone to double-mode pulsation for a
Cepheid (top) and a RR Lyrae (bottom)  model sequence;
Solid circles: stable fixed points, open circles: unstable fixed points.
($A^2$ in units of $10^{-3} (\delta R_{ph}/R_*)^2$.)}
\label{fige0e1}
\end{center}
\end{figure*}

\subsection{Amplitude Equations}

It has been known for a while now that one can develop a global understanding
of the 'modal selection' problem with the help of the amplitude equation
formalism (Buchler \& Goupil 1984).
The beauty and power of the amplitude equations are their simplicity and their
very basic and generic nature.  They apply to any system, be it a pulsating
star, a biological system, or any other dynamical system, in which (a) the
relative growth-rates of the excited modes are small; (b) the
pulsations are sufficiently weakly nonlinear so that an expansion in the
amplitudes is justified.  Only selected powers of the amplitudes appear
(Buchler \& Goupil 1984, Buchler 1993), and the omitted terms are unessential
for the modal selection (bifurcation diagram), {{\it i.e.},\ } the {\it
nature} of the dynamical behavior does not depend on them.

Condition (a) is clearly satisfied for RR~Lyrae and classical Cepheids, since
typically the growth-rates are of the order of a percent of the frequencies.
That condition (b) is also satisfied  manifests itself, for example,
by the fact that the nonlinear periods are very close to the linear ones.

The form that the amplitude equations take on depends on whether or not there
are nearby resonances. 
 \footnote{By resonance we mean a relation of the form $n_1
\omega_0 + n_2 \omega_1 \approx 0$ or $n_1 \omega_0 + n_2 \omega_1 + n_2
\omega_3 \approx 0$, where $n_1$, $n_2$, $n_3$ are small positive or negative
integers.}
  For the beat RR~Lyrae there is no low-order resonance in the period
range of interest, and for the beat Cepheids the nearest resonance is too far
to be influential.  The appropriate amplitude equations for the
nonresonant case are
\begin{eqnarray}
\dot a_0& \ngth\ngth\ngth = 
                         \ngth\ngth\ngth   
                         &a_0\th (\sigma_0 + Q_{00} \vert a_0\vert^2
                         + Q_{01} \vert a_1\vert^2
                         + S_{0} \vert a_0\vert^2 \vert a_1\vert^2\nonumber\\
                         &~&~~+ R_{00} \vert a_0\vert^4
                         + R_{01} \vert a_1\vert^4) \label{1cmplx}
\end{eqnarray}
\begin{eqnarray}						    
\dot a_1&  \ngth\ngth\ngth = \ngth\ngth\ngth  &a_1 \th (\sigma_1 
                         + Q_{10} \vert a_0\vert^2
                         + Q_{11} \vert a_1\vert^2
                         + S_{1} \vert a_0\vert^2 \vert a_1\vert^2\nonumber\\
                         &~&~~+ R_{10} \vert a_0\vert^4
                         + R_{11} \vert a_1\vert^4)
\label{2cmplx}
\end{eqnarray}
where the dot represents the time-derivative and $\sigma_j$
are the linear eigenvalues for an assumed $\exp(\sigma t)$ dependence,
\begin{equation}
\sigma_j = \kappa_j + i\omega_j
\end{equation}
The $a_j$'s are the complex amplitudes of the two excited modes, and
$Q_{jk}$, $S_j$ and $R_{jk}$ are the complex nonlinear coupling constants.

It is found to be more convenient to use real amplitudes $A_j(t)$ and phases
$\varphi_j(t)$, defined by $ a_j(t) = A_j(t) \exp i\varphi_j(t)$, because this
produces a complete decoupling of the amplitudes from the phases.  One finds
 \begin{eqnarray}
 \dot A_j  &=&  A_j\th\th\th(\kappa_j 
   + q_{j0} A_0^2 + q_{j1} A_1^2 \nonumber \\
       &\quad&\quad\quad \quad\quad 
          + s_{j} A_0^2 A_1^2 + r_{j0} A_0^4 + r_{j1} A_1^4)
 \label{aes}
 \end{eqnarray}
 and
 \begin{eqnarray}
 \dot \varphi_0 &=& \omega_0 + \hat q_{00} A_0^2
                          + \hat q_{01} A_1^2
                          + \hat s_{0} A_0^2 A_1^2 \\
 \dot \varphi_1 &=& \omega_1 + \hat q_{10} A_0^2
                          + \hat q_{11} A_1^2
                          + \hat s_{1} A_0^2 A_1^2
\end{eqnarray}

\noindent where
\begin{eqnarray}
Q_{jk}   &=& q_{jk}    +i\th \hat q_{jk}\\
S_{j}    &=& s_{j} \th\th +i\th \hat s_{j}\\
R_{jk}   &=& r_{jk}    +i\th \hat r_{jk}.
\label{qs}
\end{eqnarray}

We have disregarded the imaginary parts of the higher order terms in the phases
because they are unimportant.  One also notes that all equations can be
expressed in terms of the squares $A_0^2$ and $A_1^2$.

The constant amplitude solutions of Eqs.~(\ref{1cmplx}, \ref{2cmplx})
the so-called fixed points are therefore obtained from solving
the equations
\begin{eqnarray}
\kappa_0 &+& q_{00} A_0^2 + q_{01} A_1^2
   + s_{0} A_0^2 A_1^2  \nonumber \\ 
  &\th& \quad\quad\quad\quad + r_{00} A_0^4 + r_{01} A_1^4 =  0
  \label{ae1} \\
   A_1 & = & 0
  \label{ae2} \\  
   \nonumber \\
\kappa_1 &+& q_{10} A_0^2 + q_{11} A_1^2 
   + s_{1} A_0^2 A_1^2 \nonumber \\ 
  &\th& \quad\quad\quad\quad + r_{10} A_0^4 + s_{11} A_1^4  =  0
 \label{ae3} \\
  A_0 & = & 0 
 \label{ae4}
\end{eqnarray}
pairwise between the two sets.  
They correspond to steady nonlinear pulsations.  In the earliest use
of amplitude equations for describing DM pulsations (Buchler \& Kov\'acs 1986;
Koll\'ath {\it et al.} 1998) the amplitude equations were truncated at the
lowest nontrivial, {\it i.e.} the cubic terms.  It was shown that in this
order the single-mode fixed points and steady DM pulsations cannot
simultaneously be stable for the same stellar model.
Since then it has been realized that the behavior of the hydrodynamical models
and the observational constraints imposed by the Beat Cepheids and RR Lyrae
stars can be a little more complicated than the cubic nonresonant amplitude
equations allow.  However, the more complicated behavior can be captured if we
keep the next order nontrivial terms, \viz\ the quintic ones in the amplitudes
(Buchler {\it et al.} 1999). It is not necessary though to keep all quintic
terms, and for simplicity, these authors retained only the $r_{00}$ and
$r_{11}$ cubic terms and discussed the fixed points of the amplitude equations.
Here we keep instead the $s_0$ and $s_1$ cubic terms because they give a
slightly better description of the results of the hydrodynamical calculations.

\subsection{Determination of the Nonlinear Coefficients}
 
We can use the hydrodynamical calculations that are reported in
Fig.~\ref{figflow} to determine the values of the unknown nonlinear coupling
coefficients that appear in the amplitude equations.  This is done as follows:
The analytical signal method gives the time dependent amplitudes sufficiently
accurately so that the derivatives $\dot A$ can be used directly in
Eqs.~\ref{aes} to obtain the linear growth rates and the nonlinear coupling
coefficients in these equations through a {\sl linear} least squares fit.
(With the method described in Buchler \& Kov\'acs (1987) it was necessary to
make a tedious nonlinear fit.)

Once the coefficients are known, we can then produce a vector plot ${\bf u}
=(\dot A_0, \dot A_1)$.  The computed vector plot has been superposed on the
transient evolutionary tracks in Fig.~\ref{figflow}, and gives a global view of
the flow field.  (The dots represent the bases of the vectors).  The speeds
have been normalized because they span a very broad range of values.

The fixed point scenario is summarized in the right side of
Fig.~\ref{figflow}. The solid circles denote the stable fixed points and the
open circles the unstable ones.  We recall that the five fixed points each
correspond to a steady pulsational behavior.  The origin represents the
unstable equilibrium model (an unstable node point; for a mathematical
definition of these terms {\it cf.} Lefshetz 1977), the two fixed points on
the axes are the stable fundamental and unstable first overtone limit cycles (a
stable node point and a saddle node point, respectively).  The remaining two
fixed points are the stable and unstable double-mode pulsations (again a stable
node point and a saddle node point, respectively).  Also shown in the figure
are the integral lines that link the fixed points (the heteroclinic
connections), and the special one, the separatrix (thick line), that separates
the two attractor, \ie\ the regions of initial conditions that lead to the
stable DM pulsation and to the fundamental limit cycle, respectively.

The amplitude equations give an accurate and global picture of the
pulsational behavior of the models.

\begin{figure*}
\begin{center}
 \vspace{13pt}
 {\small
 \begin{tabular}{rrrrrrrrrrrr}
 \noalign{\smallskip}
 \noalign{\smallskip}
 \noalign{\noindent TABLE 1.\--\  
   Nonlinear Coupling Coefficients and Growth Rates.\hfill}
 \noalign{\smallskip\smallskip}
 \hline\hline
 \noalign{\smallskip}
 \hfill\ $T_{ef\!f} [K]$ \hfill
  &\hfill $q_{00}$ [$d^{-1}$]   \hfill    
  &\hfill $q_{01}$ [$d^{-1}$]   \hfill 
  &\hfill $q_{10}$ [$d^{-1}$]   \hfill 
  &\hfill $q_{11}$ [$d^{-1}$]   \hfill 
  &\hfill $s_0$ [$d^{-1}$]      \hfill 
  &\hfill $s_1$ [$d^{-1}$]      \hfill
  &\hfill $\kappa_0$ [$d^{-1}$] \hfill 
  &\hfill $\kappa_1$ [$d^{-1}$] \hfill
                           \\
 \noalign{\smallskip}
 \hline \noalign{\smallskip\smallskip\smallskip}
 \noalign{\quad Convective Cepheid Model sequence.}
 \noalign{\smallskip}
 \hline \noalign{\smallskip\smallskip\smallskip}
 \noalign{\smallskip}
  \hfill  5560  \hfill 
  &\hfill  --0.92  \hfill 
  &\hfill  --1.49  \hfill 
  &\hfill  --5.77  \hfill 
  &\hfill --11.99  \hfill 
  &\hfill  --3436.0  \hfill 
  &\hfill   2449.0  \hfill 
  &\hfill   0.00066  \hfill 
  &\hfill   0.00329  \hfill 
                     \\
  \hfill  5580  \hfill 
  &\hfill  --0.93  \hfill 
  &\hfill  --1.50  \hfill 
  &\hfill  --5.62  \hfill 
  &\hfill --12.17  \hfill 
  &\hfill  --2816.0  \hfill 
  &\hfill   4994.0  \hfill 
  &\hfill   0.00065  \hfill 
  &\hfill   0.00351  \hfill 
                     \\
  \hfill  5600  \hfill 
  &\hfill  --0.94  \hfill 
  &\hfill  --1.45  \hfill 
  &\hfill  --5.48  \hfill 
  &\hfill --12.03  \hfill 
  &\hfill  --2653.0  \hfill 
  &\hfill   6594.0  \hfill 
  &\hfill   0.00063  \hfill 
  &\hfill   0.00370  \hfill 
                     \\
  \hfill  5640  \hfill 
  &\hfill  --0.98  \hfill 
  &\hfill  --1.63  \hfill 
  &\hfill  --5.14  \hfill 
  &\hfill --11.53  \hfill 
  &\hfill  --2054.0  \hfill 
  &\hfill   5567.0  \hfill 
  &\hfill   0.00060  \hfill 
  &\hfill   0.00408  \hfill 
                     \\
  \hfill  5680  \hfill 
  &\hfill  --1.03  \hfill 
  &\hfill  --1.87  \hfill 
  &\hfill  --4.66  \hfill 
  &\hfill --10.91  \hfill 
  &\hfill  --1545.0  \hfill 
  &\hfill    830.7  \hfill 
  &\hfill   0.00057  \hfill 
  &\hfill   0.00441  \hfill 
                     \\
 \noalign{\smallskip}
 \hline \hline

 \noalign{\smallskip}
 \noalign{\smallskip}
  \noalign{\smallskip}
 \noalign{\quad Convective RR~Lyrae Model sequence.\hfill}
 \noalign{\smallskip}
 \hline \noalign{\smallskip\smallskip\smallskip}

  \hfill  6480  \hfill 
  &\hfill  --3.08  \hfill 
  &\hfill --10.28  \hfill 
  &\hfill --19.81  \hfill 
  &\hfill --44.26  \hfill 
  &\hfill  --2701.0  \hfill 
  &\hfill  --1000.0  \hfill 
  &\hfill   0.00964  \hfill 
  &\hfill   0.03953  \hfill 
                     \\
  \hfill  6500  \hfill 
  &\hfill  --3.09  \hfill 
  &\hfill --10.18  \hfill 
  &\hfill --20.24  \hfill 
  &\hfill --43.98  \hfill 
  &\hfill  --2826.0  \hfill 
  &\hfill   --545.7  \hfill 
  &\hfill   0.00936  \hfill 
  &\hfill   0.03961  \hfill 
                     \\
  \hfill  6505  \hfill 
  &\hfill  --3.07  \hfill 
  &\hfill --10.13  \hfill 
  &\hfill --20.35  \hfill 
  &\hfill --43.94  \hfill 
  &\hfill  --2750.0  \hfill 
  &\hfill   --500.4  \hfill 
  &\hfill   0.00924  \hfill 
  &\hfill   0.03967  \hfill 
                     \\
  \hfill  6515  \hfill 
  &\hfill  --3.08  \hfill 
  &\hfill --10.08  \hfill 
  &\hfill --20.62  \hfill 
  &\hfill --43.90  \hfill 
  &\hfill  --2833.0  \hfill 
  &\hfill   --383.7  \hfill 
  &\hfill   0.00910  \hfill 
  &\hfill   0.03976  \hfill 
                     \\
  \hfill  6525  \hfill 
  &\hfill  --3.07  \hfill 
  &\hfill  --9.99  \hfill 
  &\hfill --20.75  \hfill 
  &\hfill --43.72  \hfill 
  &\hfill  --2792.0  \hfill 
  &\hfill   --359.9  \hfill 
  &\hfill   0.00890  \hfill 
  &\hfill   0.03971  \hfill 
                     \\
  \hfill  6600  \hfill 
  &\hfill  --3.22  \hfill 
  &\hfill  --9.92  \hfill 
  &\hfill --22.26  \hfill 
  &\hfill --42.38  \hfill 
  &\hfill  --2781.0  \hfill 
  &\hfill     36.2  \hfill 
  &\hfill   0.00774  \hfill 
  &\hfill   0.03866  \hfill 
                     \\
  \hfill  6700  \hfill 
  &\hfill  --3.85  \hfill 
  &\hfill --10.08  \hfill 
  &\hfill --22.29  \hfill 
  &\hfill --40.16  \hfill 
  &\hfill  --2012.0  \hfill 
  &\hfill  --3303.0  \hfill 
  &\hfill   0.00605  \hfill 
  &\hfill   0.03519  \hfill 
                     \\
 \noalign{\smallskip}
 \hline\hline
 \end{tabular}
 }
 \end{center}
\end{figure*}

\section{Sequences of Models}

\subsection{Turbulent convective models}

In this section we examine constant $L$ and $M$ sequences of turbulent
convective models, one for Cepheids and one for RR ~Lyrae stars.  These
sequences display a somewhat different behavior, but that has more to do with
the chosen stellar parameters L and M than with the nature of the stars.  In
fact both Cepheids and RR Lyrae exhibit both types of behavior in some range of
L and M.

\subsubsection{Cepheid Sequence:}

In Fig.~\ref{figfpae} we report the results of the convective Cepheid model
sequence (M=4.7~{$M_{\sun} $}, L=1335~{$L_{\sun} $}) with {$T_{\rm {ef\!f}} $}\
varying from 5690 to 5530~K. 
The solid circles correspond to the amplitudes of
the computed stable DM pulsators, and the open circles to single-mode first
overtone or fundamental pulsators.

In Table~1a we show the nonlinear coupling coefficients for the Cepheid
sequence.  The calculations are laborious albeit straightforward, because for
each model in the sequence to the numerical hydrodynamics calculations have to
be made for a set of different initial conditions, and then the coefficients
have to be computed through a nonlinear fit.
One notes that the coefficients do not vary much from one model to another.
This, by the way, serves also as a confirmation of the quality of our fits.

The lines in Fig.~\ref{figfpae} show the amplitudes of the possible pulsational
states of the models.  The lines were obtained by computing the fixed points of
the amplitude equations with average values of the nonlinear coupling
constants, but using the {$T_{\rm {ef\!f}} $}\ dependent growth rates
$\kappa_0$ and $\kappa_1$.

\begin{figure*}
\resizebox{18cm}{!}{\includegraphics{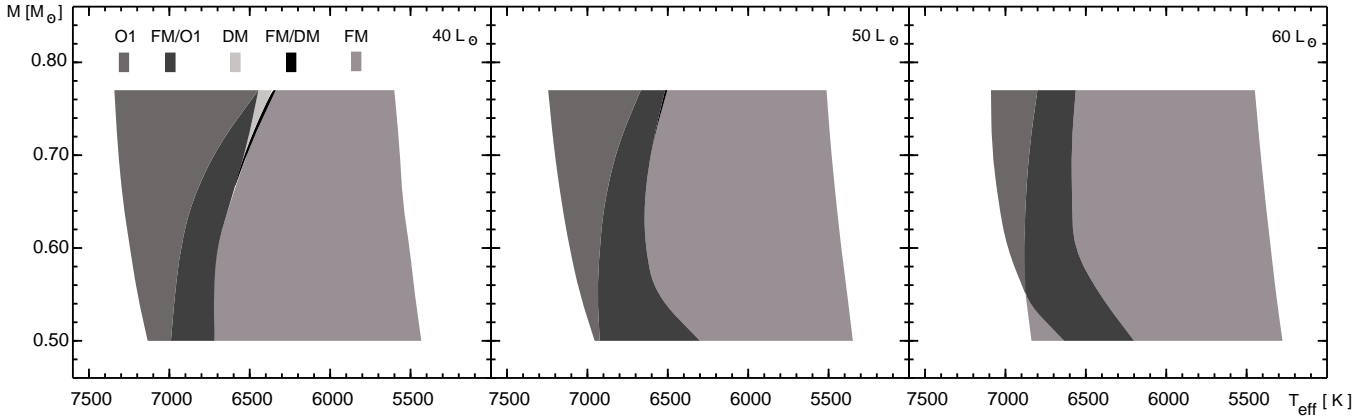}}

\caption{\small Modal selection of RR Lyrae model sequences
The gray shading of the modal states are described at 
the top of the left figure: F: fundamental mode, F/DM either 
fundamental or double mode, DM double mode, F/O1 either fundamental or 
first overtone, O1 first overtone} \label{fighrrrl}
\end{figure*}

The solid lines denote stable and the dashed lines unstable behavior.  The
first overtone limit cycle is seen to be stable down to {$T_{\rm {ef\!f}} $} =
5647~K. The fundamental limit cycle is stable up to {$T_{\rm {ef\!f}} $}=
5598~K, and a stable DM pulsation exists between {$T_{\rm {ef\!f}} $} = 5647
and 5568~K.  (A second DM pulsation is also possible between {$T_{\rm {ef\!f}}
$} = 5598 and 5568~K, but is unstable and therefore astronomically not
achievable.)  The sequence thus shows hysteresis, {{\it i.e.}} in the range
{$T_{\rm {ef\!f}} $} = 5598 to 5568~K the pulsational state depends on whether
the stellar evolution is blueward or redward.

Because Eqs.~(\ref{ae1} \th--\th \ref{ae4}) are a function of the squared
amplitudes, another visualization of the bifurcation scenario is afforded by
Fig.~\ref{fige0e1} which shows the fixed points in an ($A_0^2$,$A_1^2$) plot
for the same Cepheid model sequence.  The loci defined by Eqs.~(\ref{ae1},
\ref{ae3}) are shown as dashed and solid lines, respectively. The remaining
two loci are the positive x and y axes. The solid and open circles represent
stable and unstable fixed points which are the intersections of these loci.

\subsubsection{RR~Lyrae Sequence:}

The results of this sequence are reported in Fig.~\ref{figfpae}.  Here we have
the possibility of a fundamental limit-cycle with {$T_{\rm {ef\!f}} $}\ $<$
6700K, of an overtone limit-cycle with {$T_{\rm {ef\!f}} >$} 6500~K.  The
{$T_{\rm {ef\!f}} $}\ range of DM behavior is extremely narrow.  Again we have
hysteresis. Along a redward track the model starts off as an RRc star, then
becomes a beat RR Lyrae (RRd) and finally changes into an RRab.  Along a
blueward excursion the model starts of a RRab and then turns into an RRc,
without ever being RRd.

Compared to Fig.~\ref{figfpae} the situation is now reversed in terms of the
relative locations of points ${\bf \alpha}$ and ${\bf\gamma}$.  Again, if we
could disregard the quintic terms for the sake of argument, points ${\bf
\alpha}$ and ${\bf\beta}$ would merge, and in the scenario described by Buchler
\& Kov\'acs (1986), we are now in an 'either fundamental or overtone' situation
(${\cal D} < 0$), as confirmed by the values in Table~1b for a sequence of
RR~Lyrae (M=0.77~{$M_{\sun} $}, L=50.0~{$L_{\sun} $}, Z=0.0001)
models.  The DM behavior
occurs here only because of the looping back due to the quintic terms,
and as a result is much narrower than in the Cepheid case.

Fig.~\ref{fige0e1} shows the fixed points in an ($A_0^2$,$A_1^2$) plot for the
same RR Lyrae model sequence.  The loci defined by Eqs.~(\ref{ae1}, \ref{ae3})
are shown as dashed and solid lines, respectively.  The remaining two loci
are the positive x and y axes. The solid and open circles represent stable and
unstable fixed points which are the intersections of these loci.

We summarize the modal selection results obtained for a number of RR~Lyrae
sequences in Figure~\ref{fighrrrl} which shows the regions in which the various
pulsational behavior occurs.  In these calculations the metallicity was at
Z=0.0001, since according to our calculations the metallicity has only a minor
effect of the RR Lyrae edges.  All blue and red edges are nonlinear ones except
for the F red edge, which is from linear stability analysis.  Note that at low
luminosity and high mass an additional F region appears. We caution again that
these results depend on our choice of turbulent convective parameters and that
they may have to be revised, at least quantitatively, when a more definitive
calibration is performed.

\section{Discussion}

We have already pointed out that radiative models have failed to yield steady
DM pulsations, at least with a behavior that satisfies the observations. It is
therefore of interest to dwell a little on why convective models do.

First we note that the looping back between 5600 and 5570K in the Cepheid
models and between 6510 and 6500K for RR Lyrae models in Fig.~\ref{figfpae}
occurs because of the presence of quintic terms in the amplitude equations.  If
for the sake of argument we were to ignore the quintic terms then the
${\bf\alpha}$ and ${\bf\gamma}$ points in Fig.~\ref{figfpae} would merge, and
would be to the right of the point ${\bf\alpha}$, where the fundamental
amplitude vanishes.

\begin{figure}
\resizebox{8cm}{!}{\includegraphics{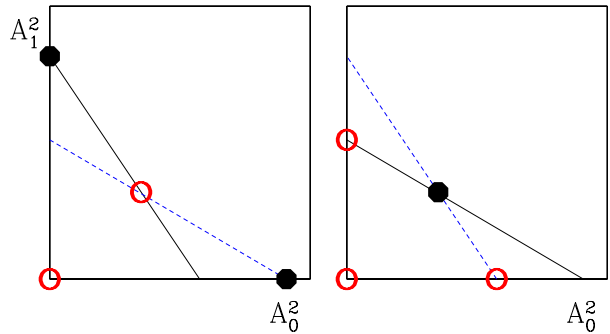}}
\caption{
Schematic plot.
Solid circles: stable fixed points, open circles: unstable fixed points.
Left: F or O1 regime, ${\cal D}<0$ and Right: DM regime, ${\cal D}>0$.}
\label{fige0e1rad}
\end{figure}

In Figs.~\ref{fige0e1} the curves would turn into straight lines, as shown
schematically in Fig.~\ref{fige0e1rad}.  In the situation depicted on the left
side the model pulsates in either the F or the O1 limit cycle, depending on the
past evolutionary history.  The right side corresponds to the DM case.  It is
easily shown (Buchler \& Kov\'acs 1986) that a {\it necessary} condition for
double-mode behavior is that ${\cal D} = q_{00} q_{11} -q_{01} q_{10} >0$.
This condition is never found to be satisfied in purely radiative models in
which the cross-coupling terms $q_{01}$ and $q_{10}$ always dominate over the
self saturation terms $q_{00}$ and $q_{11}$.

One notes in Fig.~\ref{fige0e1} that the \Teff =5610K Cepheid model has the
same topology as the right figure with the DM behavior, and that the
\Teff=6530K RR Lyrae model has that of the left side.  Table~1 shows that
indeed ${\cal D}>0$ for the Cepheid model sequence and ${\cal D}<0$ for the RR
Lyrae one, even though this criterion applies only approximately when the
quintic terms are considered.  Nevertheless it explains why the DM region in
the convective Cepheid model sequence is much broader than that of the RR
Lyrae (it would exist even in the absence of the quintic terms).


In Koll\'ath et al. (1998) we had somewhat hastily suggested that the abrupt
disappearance of the DM solution was probably due to a pole caused by the
vanishing of ${\cal D}$.  There is indeed a pole in the broad vicinity, but it
is too far to play a role.  Instead, as we have shown here, the behavior is due
to the appearance and collision of two DM fixed points, a situation that is
made possible because of the importance of the quintic terms which generate the
two DM fixed points.

\section{Conclusions}

We have shown that there are two reasons for the occurrence of DM behavior with
the turbulent convective models.  One is the increase of the cubic saturation
terms, $q_{00}$ and $q_{11}$, relative to the cross-coupling terms, $q_{01}$
and $q_{10}$.  This is a rather subtle and hidden effect of turbulent
dissipation because these coefficients are given as complicated integrals over
the structure of the star (Buchler \& Goupil 1984).

The other reason is the necessity to include quintic terms in
the description which, in turn, enrich the complexity of the bifurcation
diagram, in particular by causing the simultaneous occurrence of two DM fixed
points -- compare Figs.~\ref{fige0e1} and \ref{fige0e1rad}.

One of the interesting questions that we have not addressed concerns the time
it takes to switch from one pulsational mode (fixed point) to another one.
This problem, in our opinion, had never been correctly addressed in the
pulsation literature, and we discuss it in a companion paper (Buchler \&
Koll\'ath 2001) where we show that this transition does generally not occur on
the thermal time scale $t_{th} = \kappa^{-1}$, but on a much longer time scale,
which, depending on the type of bifurcation is either set by the stellar
evolution time scale, more precisely $(d\kappa/dt)^{-\ah}$, or by the thermal
time  $t_{th}$, however generally multiplied by several orders of magnitude.

\begin{acknowledgements}

This work has been supported by the National Science Foundation (AST9819608)
and by the Hungarian OTKA (T-026031).

\end{acknowledgements}


\begin{thebibliography}{}

\bibitem[1994]{bs}
Bono, G.\& Stellingwerf, R.F. 1994, ApJS, 93, 233

\bibitem[1993]{mito}
Buchler, J. R. 1993, in {\it Nonlinear Phenomena in Stellar
Variability}, Eds. M. Takeuti \& J.R. Buchler (Kluwer: Dordrecht), repr. from
ApSS 210, 1.

\bibitem[1984]{bg84}
Buchler, J.R. \& Goupil, M.J. 1984, ApJ, 279, 394

\bibitem[2001]{bk2001}
Buchler, J.R. \& Koll\'ath, Z. 2001, A\&A to be submitted 

\bibitem[1986]{bk86}
Buchler, J.R. \& Kov\'acs, G. 1986, ApJ, 308, 661

\bibitem[1987]{bk87}
Buchler, J.R. \& Kov\'acs, G. 1987, ApJ, 318, 232


\bibitem[1999]{bykg}
Buchler, J. R., Yecko, P., Koll\'ath, Z. \& Goupil, M. J. 1999,
{\it  ASP Conference Series} 183, 141

\bibitem[1994]{cohen}
Cohen, L. 1994, Time-Frequency Analysis. Prentice-Hall PTR. Englewood Cliffs, NJ

\bibitem[1998]{f98}
Feuchtinger, M.U 1998, A\&A, 337, 29

\bibitem[1946]{gabor}
G\'abor, D. 1946, Theory of communications, J.IEEE (London), 93, 429

\bibitem[1992]{gw92}
Gehmeyr, M. \& Winkler, K.H.A. 1992,
A\&A 253, 92; ibid. 253, 101

\bibitem[1998]{kb98}
Koll\'ath, Z., Beaulieu, J.P., Buchler, J.R. \& Yecko, P. 1998,
ApJ Letters, 502, L55
 
\bibitem[2001]{kb01}
Koll\'ath, Z. \& Buchler, J. R.  2001, in {\it Nonlinear Stellar Pulsation},
Eds. M. Takeuti \& D. Sasselov, Astrophys. \& Space Sci. Library
Vol. 257. p. 29., (astro-ph/0003386)

\bibitem[2001]{geza}
Kov\'acs, G. 2001, in {\it Nonlinear Stellar Pulsation}, Eds. M. Takeuti \&
D. Sasselov, Astrophys. \& Space Sci. Library Vol. 257.  
p. 61 (astro-ph/0003386) 

\bibitem[1993]{kb}
Kov\'acs, G. \& Buchler, J.R. 1993, ApJ, 404, 765

\bibitem[1987]{kbd87}
Kov\'acs, G., Buchler, J.R. \& Davis, C.G. 1987, ApJ, 319, 247

\bibitem[1986]{kuhfuss}
Kuhfuss, R. 1986,
A\&A, 160, 116

\bibitem[1977]{lefshetz}
Lefshetz, S. 1977, {\it Differential Equations: Geometric Theory}, 
(New York: Dover)


\bibitem[1982]{stellingwerf82}
Stellingwerf, R.F. 1982,
ApJ, 262, 330

\bibitem[1998]{wf}

Wuchterl, G. \& Feuchtinger, M.U. 1998
A\&A, 340, 419

\bibitem[1998]{ykb}
Yecko, P., Koll\'ath Z. \& Buchler, J.R. 1998, A\&A, 336, 553

\end{thebibliography}
\end{document}